\documentclass[prd,aps,preprintnumbers,fleqn,showpacs,nofootinbib,superscriptaddress]{revtex4}

\usepackage{amssymb,amsmath,graphicx,bm}

\usepackage[usenames,dvipsnames]{color}
\usepackage[normalem]{ulem} 
\renewcommand\sout{\bgroup \color[rgb]{0.55,0.00,0.99} \ULdepth=-.5ex \ULset}

\def\slash#1{\setbox0=\hbox{$#1$}               
        \dimen0=\wd0                            
        \setbox1=\hbox{/} \dimen1=\wd1          
        \ifdim\dimen0>\dimen1                   
        \rlap{\hbox to \dimen0{\hfil/\hfil}}    
        #1                                      
        \else
        \rlap{\hbox to \dimen1{\hfil$#1$\hfil}} 
        /                                       
        \fi}                                    %

\newcommand{\beqn}{\begin{eqnarray}}
\newcommand{\eeqn}{\end{eqnarray}}

\begin{document}

\title{Impact parameter dependence of the azimuthal asymmetry in  lepton pair production in
 heavy ion collisions}

\author{Cong Li}
 \affiliation{\normalsize\it Key Laboratory of
Particle Physics and Particle Irradiation (MOE),Institute of
frontier and interdisciplinary science, Shandong University(QingDao), Shandong 266237, China }

\author{Jian~Zhou}
 \affiliation{\normalsize\it Key Laboratory of
Particle Physics and Particle Irradiation (MOE),Institute of
frontier and interdisciplinary science, Shandong University(QingDao), Shandong 266237, China }

\author{Ya-jin Zhou}
\affiliation{\normalsize\it Key Laboratory of Particle Physics and
Particle Irradiation (MOE),Institute of frontier and
interdisciplinary science, Shandong University(QingDao), Shandong
266237, China }
\begin{abstract}
We investigate  the impact parameter dependence of the $\cos 4\phi$ azimuthal asymmetry
for electromagnetic lepton pair production in heavy ion collisions. The
asymmetry induced by linearly polarized coherent  photons
exhibits strong impact parameter dependence.
\end{abstract}

\date{\today}

\maketitle

\section{Introduction}
It has long been  recognized that dilepton production in heavy ion collisions can serve as
 a clean and powerful probe for studying the electromagnetic properties of Quark Gluon
 Plasma(QGP). At  low lepton pair transverse momentum($q_\perp$), the dominant production mechanism
 is the photon-photon fusion process($\gamma \gamma\rightarrow l^+ l^-$) which is greatly enhanced
  due to the large flux of quasi-real photons coherently generated by a fast moving nucleus.
  The measurements carried out at SPS, RHIC and LHC~\cite{Agakishiev:1995xb,Arnaldi:2006jq,Adams:2004rz,Acharya:2018nxm,Aaboud:2018eph,Adam:2018tdm,Adam:2018qev}
   have clearly demonstrated that dilepton production is indeed peaked at very low transverse momentum.
 The comprehensive understanding of such process is not only important for accurate background estimates
 in heavy ion collisions, but also interesting from a pure theoretical point of view. For example,
 there is so far no clear experimental evidence for  Coulomb correction
 effects~\cite{Bethe:1954zz,Davies:1954zz,Segev:1997yz,Ivanov:1998ka,Baltz:2001dp,Eichmann:1998eh,Baltz:2007gs}.
  Moreover, whether the significant $ q_\perp^2$ broadening effects   in peripheral and central collisions
 recently observed by ATLAS and STAR collaborations
  results from the initial state effect or the various final state effects is still under
 debate~\cite{Klein:2018cjh,Zha:2018ywo,Klein:2018fmp,Zha:2018tlq,Ye:2018jwq}.

Electromagnetic lepton pair production  is usually computed using the equivalent photon approximation introduced
by Enrico Fermi(also often referred to as the Weizs$\ddot{a}$cker-Williams method), in which the photon flux
 is calculated by treating the fields of charged relativistic  heavy ion as external, i.e, classical electromagnetic field.
 This method has been widely used to compute Ultra-Peripheral heavy ion Collisions(UPC) observables~\cite{Bertulani:1987tz,Bertulani:2005ru,Baur:2007fv,Baltz:2007kq,Klein:2016yzr,Ma:2018zzq}.
 In the kinematic regions typically accessible at RHIC and LHC,
 these associated quasi-real photons carry  very small longitudinal momentum fraction($x$) of each nucleon inside the nucleus.
 In the small $x$ limit, there is a strong correlation between the photon's polarization tensor and
 its transverse momentum.
 In the TMD factorization framework, such correlation is characterized
 by the linearly polarized photon TMD~\cite{Mulders:2000sh}.
 It can give rise to a $\cos 4\phi$ azimuthal modulation in the dilepton
 production cross section, where $\phi$ is the angle between the lepton pair transverse momentum and
 the  individual lepton transverse momentum.
 We recently computed this $\cos 4\phi$ asymmetry and found that it is rather sizable~\cite{Li:2019yzy}.
  We note that the same observable has been studied in the context of nucleon-nucleon collisions in an
  earlier work~\cite{Pisano:2013cya}.

As a matter of fact, the similar phenomena in the QCD case has been intensively investigated in recent
years~\cite{Metz:2011wb,Dominguez:2011br,Boer:2009nc,Boer:2010zf,Qiu:2011ai,Schafer:2012yx,Pisano:2013cya,Akcakaya:2012si,Dumitru:2015gaa,Kotko:2015ura,Boer:2017xpy,Marquet:2017xwy,Gutierrez-Reyes:2019rug,Scarpa:2019fol}.
In the dilute limit, small $x$ gluons are almost fully linearly polarized just like the  Weizs$\ddot{a}$cker-Williams photons.
However, in the saturation regime,
the linear polarization of small $x$ gluons will be affected by  multiple re-scattering
 effect and thus becomes process dependent.
 It has been proposed
 to probe the linearly polarized gluon distribution by measuring $\cos 2\phi$ azimuthal asymmetries
  for two particle production in various high energy scattering processes at RHIC, LHC, or
a future Electron-Ion Collider(EIC).
 Comparing the experimental measurements of gluon polarization and photon polarization in the small $x$ limit
 would be beneficial for studying saturation effect, which is absent
 in the QED case.

In our previous work~\cite{Li:2019yzy}, we addressed the azimuthal asymmetries for dilepton production
with the equivalent photon approximation, which, however does not provide any information
about the location of the particle production process. It does not allow us to distinguish
UPC from central(or peripheral) collisions,
 as the impact parameter($b_\perp$) is implicitly integrated out in this method. The
 formalism for computing $b_\perp$ dependent cross section was developed
 in the early nineties of the past century in Refs.~\cite{Vidovic:1992ik,Hencken:1993cf}.
Calculations based on this more rigorous treatment, in some cases yield
results significantly different from that computed in the equivalent photon approximation.
 For instance, theoretical calculations  can not
 account for  the measured lepton pair transverse momentum spectrum at low $q_\perp$
   unless  $b_\perp$ dependence is taken into account\cite{Hencken:2004td,Baltz:2006mz}.

The impact parameter dependent cross section for lepton pair production
 has been calculated only for the azimuthal angle averaged case in the aforementioned literatures.
 The purpose of the present work is to investigate the impact parameter dependence of the
 $\cos 4\phi$ azimuthal asymmetry. The paper is structured as follows. In the next section, we
 derive the  $b_\perp$ dependent polarized cross section for dilepton production in heavy ion collisions.
 We present the numerical estimations for the asymmetries in the central, peripheral, and
 ultra-peripheral collisions at RHIC and LHC energy.
  The QED resummation effect is also included in our evaluations. The paper is
 summarized in Sec.III.

\section{The impact parameter dependence of  $\cos 4\phi$ azimuthal asymmetries }
The dominant channel for dilepton production
in the kinematical region where lepton pair transverse momentum
 is the order of the reverse of nucleus radius, is photon-photon fusion process,
\begin{eqnarray}
\gamma(x_1P+k_{1\perp})+\gamma(x_2 \bar P+k_{2\perp}) \rightarrow l^+(p_1)+ l^-(p_2)
\end{eqnarray}
The leptons are produced nearly back-to-back in azimuthal with  total transverse momentum
$q_\perp\equiv p_{1\perp}+p_{2\perp}=k_{1\perp}+k_{2\perp}$
being much smaller than  $P_\perp=(p_{1\perp}-p_{2\perp})/2$.
 When $P_\perp$ is sufficiently large, dilepton can be viewed as being produced locally.
 Since the location where
 dilepton is produced in the transverse plane is specified during our calculation,
 the incoming photons are no longer in the eigenstate of transverse momenta.
 Two incoming photon's momenta in the conjugate amplitude are denoted as
 $x_1P+k_{1\perp}'$ and $x_2 \bar P+k_{2\perp}'$ respectively, with the constraint
 $ k_{1\perp}'+k_{2\perp}'\equiv q_\perp$.

Following the method outlined in Refs.~\cite{Vidovic:1992ik,Hencken:1993cf}, the impact parameter dependent cross section computed at the lowest order QED reads,
 \begin{eqnarray}
\frac{d\sigma_{\!_0}}{d^2 p_{1\perp} d^2 p_{2\perp} dy_1 dy_2 d^2 b_\perp }= \frac{2\alpha_e^2}{Q^4}\frac{1}{(2\pi)^2}
\left [ \mathcal{A}+ \mathcal{B} \cos 2\phi+\mathcal{C} \cos 4\phi \right ]
\end{eqnarray}
where $\phi$ is the angle between transverse momenta $q_\perp$ and
$P_\perp$. $y_1$ and $y_2$ are leptons rapidities, respectively.
$b_\perp$ is the transverse distance between two colliding nuclei. Q is the
invariant mass of the lepton pair. The $\mathcal{B}$ term is
proportional to the lepton mass and thus very small.
 We do not present the detailed expression for $\mathcal{B}$ here.
 Instead, we focus on investigating the $\cos 4\phi$ asymmetry in this work
 which depends only on the  $\mathcal{A}$ and $\mathcal{C}$ terms.
The coefficients $\mathcal{A}$ and $\mathcal{C}$ take form,
 \begin{eqnarray}
\mathcal{A}&=& \frac{Q^2-2 P_\perp^2}{P_\perp^2}\frac{Z^4
\alpha_e^2}{\pi^4}\int d^2k_{1\perp} d^2 k_{2\perp} d^2
\Delta_\perp \delta^2( q_\perp-k_{1\perp}-k_{2\perp}) e^{i
\Delta_\perp \cdot b_\perp}
\nonumber \\&& \ \ \ \ \ \ \ \ \ \ \ \ \ \ \ \ \ \ \times \left [
(k_{1\perp} \cdot k_{1\perp}')(k_{2\perp} \cdot k_{2\perp}')+
(k_{1\perp} \! \cdot k_{2 \perp})\Delta_\perp^2
-(k_{1\perp}\!\cdot \Delta_{ \perp})(k_{2\perp}\!\cdot \Delta_{ \perp})
\right ]
\nonumber \\&& \ \ \ \ \ \ \ \ \ \ \ \ \ \ \ \ \ \ \times \
{\cal F}(x_1,k_{1\perp}^2){\cal F}^*(x_1,k_{1\perp}'^2){\cal F}(x_2,k_{2\perp}^2){\cal F}^*(x_2,k_{2\perp}'^2)
\end{eqnarray}
and
 \begin{eqnarray}
\mathcal{C} =
 -2 \frac{Z^4\alpha_e^2}{\pi^4}
\!\!\!&&\!\! \!\!\!\!  \int d^2k_{1\perp} d^2 k_{2\perp} d^2\Delta_\perp
\delta^2( q_\perp-k_{1\perp}-k_{2\perp})  e^{i\Delta_\perp \cdot b_\perp}
\nonumber \\
&\times& \left \{2\left [ 2( k_{2\perp} \! \cdot \hat q_{\perp})( k_{1\perp} \! \cdot \hat q_{\perp})
- k_{1\perp} \!\cdot \! k_{2\perp} \! \right ]
\left [ 2( k_{2\perp}' \! \cdot \hat q_{\perp})( k_{1\perp}' \! \cdot \hat q_{\perp})
- k_{1\perp}' \!\cdot \! k_{2\perp}' \! \right ] \right .\
\nonumber \\ && \left .\ -\left [
(k_{1\perp} \cdot k_{1\perp}')(k_{2\perp} \cdot k_{2\perp}')+
(k_{1\perp} \! \cdot k_{2 \perp})\Delta_\perp^2
-(k_{1\perp}\!\cdot \Delta_{ \perp})(k_{2\perp}\!\cdot \Delta_{ \perp})
\right ]\right \}
\nonumber \\
& \times &
{\cal F}(x_1,k_{1\perp}^2){\cal F}^*(x_1,k_{1\perp}'^2){\cal F}(x_2,k_{2\perp}^2){\cal F}^*(x_2,k_{2\perp}'^2)
\end{eqnarray}
where $\Delta_\perp=k_{1\perp}-k_{1\perp}'=k_{2\perp}'-k_{2\perp}$.
 $\hat q_\perp$ is unit vector defined as  $\hat q_\perp= q_\perp/| q_\perp| $.
The incoming photons  longitudinal momenta fraction  are fixed by the
external kinematics according to   $x_1=\sqrt{\frac{P_\perp^2+m^2}{s}}(e^{y_1}+e^{y_2})$
 and $x_2=\sqrt{\frac{P_\perp^2+m^2}{s}}(e^{-y_1}+e^{-y_2})$ with $m$ being the lepton mass and $s$ being
 the center mass energy.
 The nuclear charge form factor enters the cross section via
${\cal F}(x,k_\perp^2)=\frac{F(k_{\perp}^2+x^2M_p^2)}{(k_{\perp}^2+x^2M_p^2)}$,
 where $M_p$ is proton mass. Note that the lepton mass is ignored in the hard coefficients.

In comparison with our previous work where the calculation is formulated in TMD factorization,
 here we directly express the cross section as the convolutions of the form factor. Alternatively, one can
 factorize  the impact parameter dependent cross section in terms of the photon Wigner distribution.
 However, this is beyond the scope of the present paper and will be addressed in a separate publication.
 Of course, if $b_\perp$ is integrated out in the above cross section formula, one can recover the cross section~\cite{Li:2019yzy}
 derived using the equivalent photon approximation. This provides a nice consistency check.
\begin{figure}[htpb]
\includegraphics[angle=0,scale=1.1]{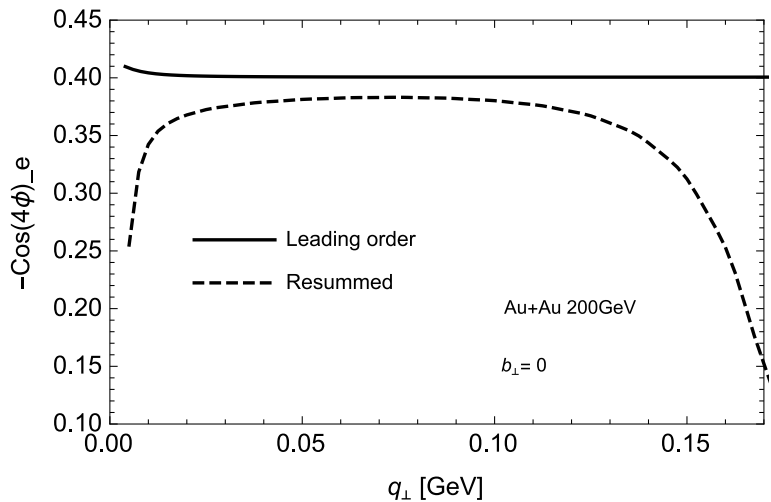}
\caption{ The $\cos 4\phi$ asymmetry in Au-Au collisions at the impact parameter $b_\perp=0$
 with and without taking into account resummation effect.
The center mass energy is  $\sqrt {s}=200 $ GeV.
The electron and positron rapidities and transverse momenta are integrated over the regions [-1,1],
 and [0.2 GeV, 0.4 GeV] respectively.} \label{fig1}
\end{figure}
The $\cos 4\phi$ azimuthal asymmetry is determined by the ratio between the terms $\mathcal{A}$ and $\mathcal{C}$.
 In general, the size of the asymmetry  can only be numerically calculated because of the complicated
 convolutions  involved. However, quite remarkably, the analytical solution is available at
 $b_\perp=0$. After a few steps of algebraic manipulations,  one finds that
 the convolutions in $\mathcal{A}$ and $\mathcal{C}$ terms turn out to be
identical for $b_\perp=0$. The asymmetry is then simply proportional to,
\begin{eqnarray}
\frac{\mathcal{C}(b_\perp=0)}{2\mathcal{A}(b_\perp=0)}= \frac{-2P_\perp^2}{2(Q^2-2 P_\perp^2)}
\end{eqnarray}
which indicates that the $\cos 4\phi$ asymmetry for pure electromagnetic lepton pair production
 in the central collisions is independent of $q_\perp$.
 This finding has been verified by the explicit numerical estimation as shown in Fig.1.
 However, one should keep in mind that hadronic background contribution in central collisions is significant.
The size of the measured asymmetry could be far below the predicated value.

We now proceed to describe the ingredients needed in the numerical evaluations of the asymmetry.
 First,  the form factor is taken
 from the STARlight MC generator~\cite{Klein:2016yzr},
\begin{eqnarray}
F(|\vec k|)=\frac{4\pi \rho^0}{|\vec k|^3 A}\left [ \sin(|\vec k|R_A)-|\vec k|R_A \cos(|\vec k|R_A)\right ]\frac{1}{a^2 \vec k^2+1}
\label{ff}
\end{eqnarray}
where   $a=0.7$ fm, and $\rho^0$ is a normalization factor. The nucleus radius is chosen to be $R_A=1.1 A^{1/3}$fm for Au and Pb targets,
 $R_A=1.2 A^{1/3}$fm for Ru target.
This parametrization is very close to the Woods-Saxon distribution,
and is used in our numerical evaluation. In order to take into account the effect of final state multiple
soft photon radiation,
 a Sudakov factor has to be inserted in the differential cross section in  $r_\perp$ space,
\begin{eqnarray}
\frac{d\sigma}{d^2 p_{1\perp} d^2 p_{2\perp} dy_1 dy_2 d^2 b_\perp }= \int
\frac{d^2 r_\perp}{(2\pi)^2} e^{i r_\perp \cdot q_\perp} e^{- S(Q,r_\perp)} \int d^2 q_\perp'
e^{i r_\perp \cdot q_\perp'} d\sigma_{_{\!0}}(q_\perp',  \ ...)
\end{eqnarray}
where the Sudakov factor at one loop is given by~\cite{Klein:2018fmp},
\begin{eqnarray}
S(Q,r_\perp)=
\left \{
\begin{aligned}
&& \frac{\alpha_e}{2\pi} {\rm ln}^2 \frac{Q^2}{\mu_r^2}, \ \ \ \ \ \  \mu_r>m_\mu \\
&& \frac{\alpha_e}{2 \pi} {\rm ln}\frac{Q^2}{m_\mu^2} \left [
{\rm ln}\frac{Q^2}{\mu_r^2} +{\rm ln} \frac{m_\mu^2}{\mu_r^2}
\right ], \ \ \ \ \ \ \mu_r<m_\mu
\end{aligned}
\right.
\end{eqnarray}
with $\mu_r=2 e^{-\gamma_E}/|r_\perp|$.
The perturbative tail at relatively high $q_\perp$ is mainly generated by this Sudakov factor.
 The theoretical calculation is in good agreement with the ATLAS UPC high $q_\perp$ data~\cite{Klein:2018fmp}.
In contrast, at intermediate lepton pair transverse momentum,
roughly speaking  $10 \ \text{MeV} <q_\perp<70$ MeV, the $q_\perp$ shape is determined by the primordial
photon distribution.  At low $q_\perp$($<$10MeV), the smallness of
 the fine coupling constant is compensated by the large double logarithm ${\rm ln}^2 \frac{Q^2}{\mu_b^2}$, leading to
 a significant resummation effect, which, for example, has been clearly demonstrated in Fig.1.

\begin{figure}[htpb]
\includegraphics[angle=0,scale=0.9]{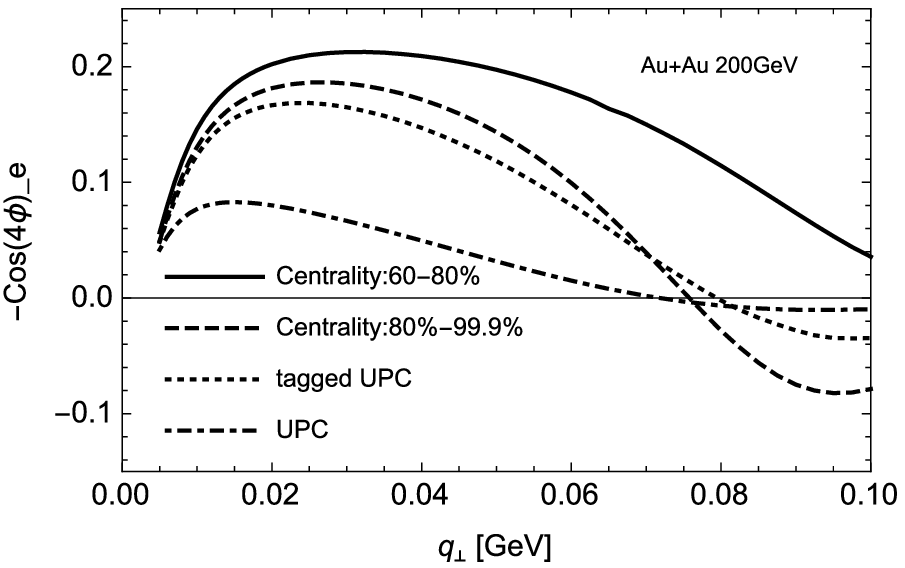}
\includegraphics[angle=0,scale=0.9]{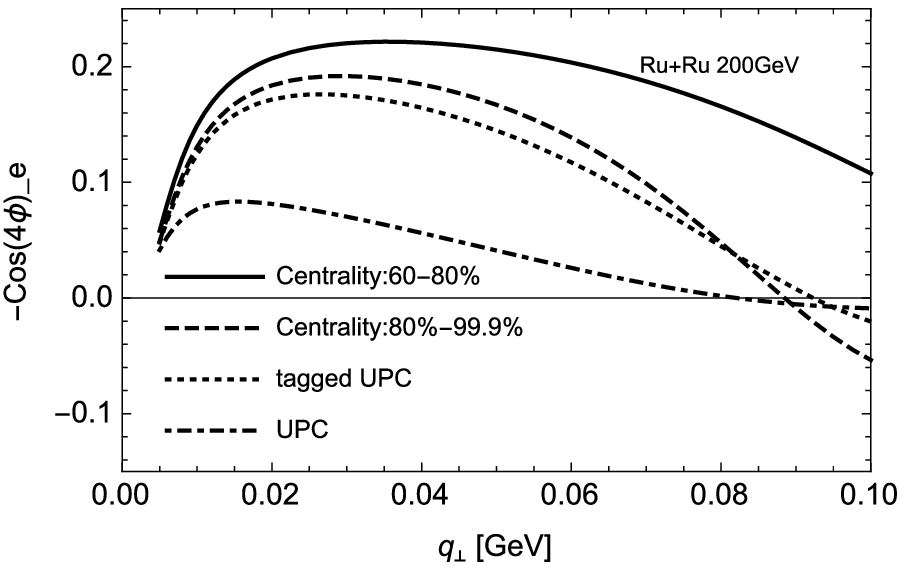}
\caption{ Estimates of the $\cos 4\phi$ asymmetry as the function of $q_\perp$
 for the different  centralities at  $\sqrt {s}=200 $ GeV.
The electron and positron  rapidities and transverse momenta are integrated over the regions [-1,1],
 and [0.2 GeV, 0.4 GeV]. The asymmetries in Au-Au collisions and Ru-Ru collisions
 are shown in the left plot and the right plot respectively.
 } \label{fig2}
\end{figure}

\begin{figure}[htpb]
\includegraphics[angle=0,scale=1.1]{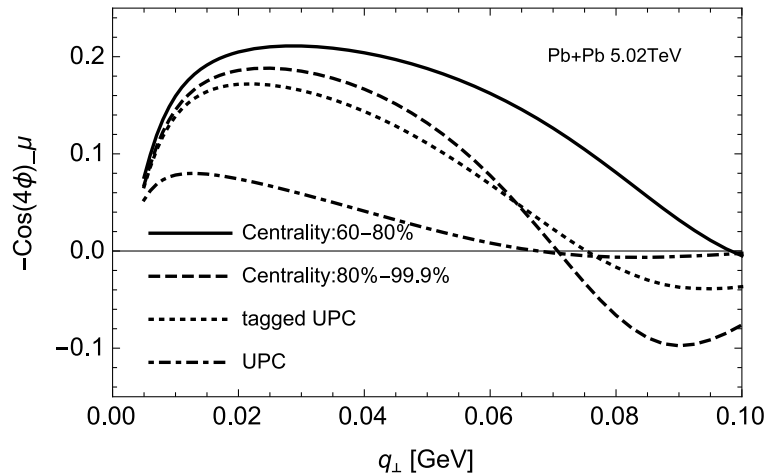}
\caption{ Estimates of the $\cos 4\phi$ asymmetry as the function of $q_\perp$
 for the different  centralities at  $\sqrt {s}=5020 $ GeV.
The muon and anti-muon rapidities and transverse momenta are integrated over the regions [-1,1],
 and [4 GeV, 45 GeV] respectively.
 } \label{fig3}
\end{figure}

 The numerical results for the computed azimuthal asymmetries for the different collisions species and
 centralities are
 presented in Figs.\ref{fig2} and \ref{fig3}. Here the azimuthal asymmetries, i.e. the average value of  $\cos 4\phi$ are defined as,
\begin{eqnarray}
\langle \cos(4\phi) \rangle &=&\frac{ \int \frac{d \sigma}{d {\cal P.S.}} \cos 4\phi \ d {\cal P.S.} }
{\int \frac{d \sigma}{d {\cal P.S.}}  d {\cal P.S.}}
\end{eqnarray}
We compute the asymmetry for two deferent centrality classes as well as for the UPC and the tagged UPC cases.
 The corresponding  impact parameter range for a given centrality class is determined
  using the Glauber model(see the review article~\cite{Miller:2007ri} and references therein).
    For the UPC,  the asymmetry is averaged over the impact parameter range $[2R_A, \infty]$.
  However, STAR experiments at RHIC measure pair production cross section
  together with the double electromagnetic excitation in both ions. Neutrons
  emitted at forward angles by the fragmenting nuclei are measured, and used as a UPC trigger.
 Requiring lepton pair to be produced  in coincidence with Coulomb breakup of the beam nuclei
 alters the impact parameter distribution compared with exclusive production.
 In order to incorporate the experimental conditions in the theoretical
 calculations, one can define a "tagged" UPC cross section,
\begin{eqnarray}
2 \pi \int_{2R_A}^{\infty} b_\perp db_\perp P^2(b_\perp) d \sigma(b_\perp, \ ...)
\end{eqnarray}
 where the probability $P(b_\perp)$ of emitting
 a neutron from  the scattered nucleus is often parameterized as~\cite{Baur:1998ay},
\begin{eqnarray}
P(b_\perp)= 5.45*10^{-5}\frac{Z^3(A-Z) }{A^{2/3} b_\perp^2}\exp \left [-5.45*10^{-5}\frac{Z^3(A-Z) }{A^{2/3} b_\perp^2} \right ]
\end{eqnarray}
  As a matter of fact, the
 mean impact parameter is dramatically reduced in interactions with  Coulomb dissociation.

  We plot the $\cos 4\phi$ asymmetry for electron pair production
at mid-rapidity as the function of the total transverse momentum $q_\perp$
at the center mass energy $\sqrt s=200$ GeV in Fig.\ref{fig2}. The general trend is that
 the asymmetry increases  when the impact parameter decreases.
 The overall $q_\perp$ and $b_\perp$ dependent behavior of the
 asymmetry for the different collision species(Au and Ru) are similar, except for that the curves are slightly
 more flat for the smaller nucleus.
 The asymmetry reaches a maximal value of 17\%--22\% percent around $q_\perp\approx 30$ MeV for the centrality
 classes [60\%-80\%], [80\%-99.9\%], and the tagged UPC. For the unrestricted UPC, the asymmetry is roughly
 twice smaller than that in the tagged UPC. The results obtained for di-muon production in
 Pb-Pb collisions at LHC energy shown in Fig.\ref{fig3} are rather close to these at RHIC energy.

\section{Conclusions}
We study the impact parameter dependence of the $\cos 4\phi$ azimuthal asymmetry for purely electromagnetic
 lepton pair production in heavy ion collisions at low $q_\perp$.
 This asymmetry arises from the correlation between
  the  polarization vector of the electric field coherently generated by a fast moving heavy ion
   and the associated equivalent photon's transverse momentum.
   Such correlation reflects the nature of the boosted Coulomb potential.  We found that the azimuthal asymmetry
   has a  strong $b_\perp$ dependence. To be more specific, the asymmetry
 decreases with increasing impact parameter. Moreover, the $q_\perp$ dependent behavior of the
 azimuthal asymmetry is different  in the different $b_\perp$ regions.
  We present numerical results for the $b_\perp$ and $q_\perp$
    dependent asymmetry  for the different collision species at various center mass energies.
 It would be interesting to test these theoretical predications  at RHIC and LHC.

 The study of such initial state effect in heavy ion collisions is not only important for facilitating
 the investigations of the electromagnetic properties of QGP,
  but also interesting in its own right. For instance, the polarization dependent
 observable can be used as a powerful tool to study QED processes in strong electromagnetic fields,
 particularly in view of the fact that no definitive conclusion on Coulomb correction
  has yet been drawn on experimental side.
 It has an advantage over the azimuthal angle averaged cross section in this regard  because the absolute normalization
  which suffers various uncertainties are cancelled out when computing the asymmetry.
 Furthermore,  comparing photon's linear polarization and gluon's linear polarization in the small $x$ limit
  would be  helpful for us to gain more insight into how the gluon polarization is affected by saturation effect.

\begin{acknowledgments}
We thank Zhang-bu Xu,  Wang-mei Zha, and James Daniel Brandenburg for drawing
 our attention to the Ref.\cite{Vidovic:1992ik}.
  J. Zhou thanks  Feng Yuan, Bowen Xiao, and Chi Yang  for helpful discussions.
 J. Zhou has been supported by the National Science Foundations of
China under Grant No.\ 11675093, and by the Thousand Talents Plan
for Young Professionals. Ya-jin Zhou has been supported by the
National Science Foundations of China under Grant
No.\ 11675092.
\end{acknowledgments}

\end{document}